\documentclass[aps,prl,floatfix,reprint,showpacs,superscriptaddress]{revtex4-1}
\usepackage[utf8x]{inputenc}
\usepackage{graphicx,epsfig,color}

\begin{document}

\title{Re-assigning (1$\times$2) reconstruction of rutile TiO$_\mathbf{2}$(110)
from DFT+\emph{U} calculations}

\author{Hatice \"{U}nal}
\author{Ersen Mete}\email{emete@balikesir.edu.tr}
\affiliation{Deparment of Physics, Bal{\i}kesir University, Bal{\i}kesir 10145,
Turkey}
\author{\c{S}inasi Ellialt{\i}o\u{g}lu}
\affiliation{Department of Physics, Middle East Technical University, Ankara
06800, Turkey}

\date{\today}

\begin{abstract}
Physically reasonable electronic structures of reconstructed rutile
TiO$_2$(110)-(1$\times$2) surfaces were studied using density functional
theory (DFT) supplemented with Hubbard $U$ on-site Coulomb repulsion acting
on the $d$ electrons, so called as the DFT+$U$ approach. Two leading
reconstruction models proposed by Onishi--Iwasawa and Park \textit{et al.} were
compared in terms of their thermodynamic stabilities.
\end{abstract}


\maketitle

\begin{figure}[tbh]
\epsfig{file=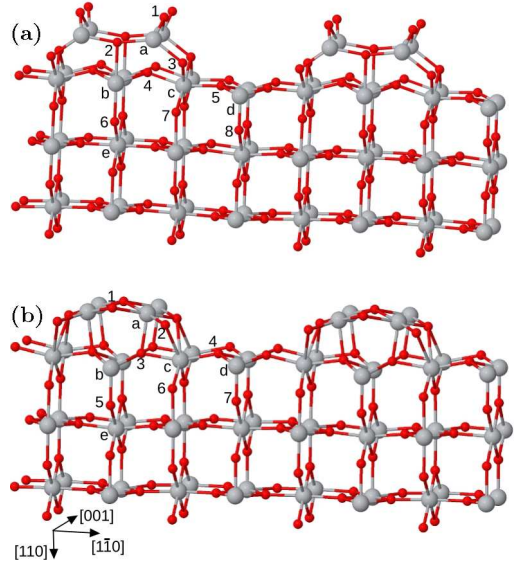,width=6.7cm}
\caption{Fully relaxed (1$\times$2) reconstructed rutile TiO$_2$(110)
surface models of (a) Onishi--Iwasawa~\cite{onishi} and (b) Park
~\textit{et al.}~\cite{park} showing 3 Ti-layers from the top.\label{fig1}}
\end{figure}

Rutile TiO$_2$ and its surfaces represent model systems to explore the
properties of transition metal oxides that are important in technological
applications such as catalysis, photovoltaics, and gas sensing~\cite{diebold1},
to name a few. Truncated or stoichiometric (110) surface of rutile is the most
stable one among all surfaces of titania~\cite{heinrich}. Upon thermal annealing
or ion bombardment TiO$_2$(110)-(1$\times$1) surface is reduced by loosing the
bridging oxygens, and is often undergo a (1$\times$2) reconstruction with row
formations~\cite{onishi,murray,guo,pang,elliot1,elliot2,mccarty,blancorey1,
blancorey2,park, shibata}. The identification of these rows on reconstructed
surfaces, in three dimensions, is difficult by experimental methods~\cite{shibata}. 
The best candidate for modeling this reconstruction involves
the addition of ``Ti$_2$O$_3$" molecule on the surface unit cell (added-row
model) proposed by Onishi and Iwasawa~\cite{onishi}. In addition to theoretical
studies it was supported by electron stimulated desorption of ion angular
distribution (ESDIAD), scanning tunnelling microscopy (STM), and low-energy
electron diffraction (LEED) experiments~\cite{guo,pang,elliot1,elliot2,mccarty,
blancorey1,blancorey2}. On the other hand, Shibata~\textit{et al.}~\cite{shibata},
using advanced transmission electron microscopy (TEM) observations reported
results that are consistent with a different model, proposed by Park \textit{et
al.}~\cite{park}, for which the additional unit is ``Ti$_2$O". The main
difference between the Onishi--Iwasawa and Park~\textit{et al.} models (Onishi
and Park models, respectively, in what follows) is the locations of Ti
interstitial sites on the surface~\cite{shibata} (see Fig.~\ref{fig1}).

Now there is a debate about which formation gives rise to the (1$\times$2)
long range order, the last proposed one or the best previous candidate? In
this context we study the electronic properties of these two models using
Hubbard $U$ corrected total energy density functional theory (DFT+$U$)
calculations to get physically reasonable results comparable to existing
experimental data. We discuss which of these leading models can be assigned to
describe the (1$\times$2) reconstructed surface by comparing them according to
their thermodynamic stabilities.

Band structures of reduced and reconstructed TiO$_2$(110) surfaces determined by
pure DFT calculations does not agree with experiments~\cite{kimmel,kruger,
blancorey1}. Failure of the standart DFT is not limited to band-gap
underestimation stemming from the many-electron self-interaction error (SIE).
More importantly, it does not predict experimentally observed gap states
\cite{henrich, henderson} that are associated with the excess electrons
due to the formation of surface oxygen vacancies. In DFT calculations, these
Ti $3d$ electrons occupy the bottom of the conduction band (CB) giving metallic
character. Hence, hybrid DFT methods need to be used. For instance, SIE can be
partly corrected by partially mixing nonlocal Fock exchange term with DFT
exchange term~\cite{yfzhang,valentin} or DFT+$U$ approach~\cite{dudarev} can
make up for the lack of strong correlation between the 3$d$ electrons, a
shortcoming of common exchange--correlation functionals. Empirical Hubbard $U$
term accounts for the on-site Coulomb repulsion between the Ti $3d$ electrons.
By examining different values of the $U$ parameter, experimentally
observed gap state of the reduced TiO$_2$(110) surface was obtained
$\sim$0.7--0.9 eV below the CB~\cite{morgan,calzado,nolan}.

For the (1$\times$2) reconstructed surface with Ti$_2$O$_3$ added row,
Kimura~\textit{et al.}~\cite{kimura}, and then, Blanco-Rey~\textit{et al.}
\cite{blancorey1}, obtained the Ti 3$d$ states positioned inside the CB, and
proposed this model to be metallic by pure DFT methods. Recently, using spin
polarized DFT+$U$ calculations with a suitable choice of the $U$ parameter, we
have shown that the Onishi model can be semiconducting~\cite{celik} as
the reconstructed surface is observed experimentally~\cite{abad}. Important
questions still remain to be answered such as where the excess charge
density is distributed and which model structure describes the (1$\times$2)
reconstruction.


We used Perdew--Burke--Ernzerhof (PBE)~\cite{pbe} gradient corrected
exchange--correlation functional supplemented with Dudarev's $U$
term~\cite{dudarev} as implemented in the VASP code~\cite{vasp}.
The ionic cores and valence electrons were treated by the projector-augmented
waves (PAW) method~\cite{paw1,paw2} up to a cutoff value of 400 eV. In order
to get well converged energetics, we adopted stoichiometric slabs with 10
Ti-layers (30 atomic layers) separated by $\sim$15 {\AA} vacuum from their
periodic images. We built the Onishi and the Park models of (1$\times$2)
reconstructed rutile (110) surfaces by adding Ti$_2$O$_3$ and Ti$_2$O groups,
respectively, along [001] both at the top and at the bottom of the slabs (see
Fig.~\ref{fig1}). Hence, having an unphysical dipole across the slab and getting
two different groups of surface states in the gap has been avoided. Instead,
symmetric slabs bring about the same group of surface states (degenerate) from
both surfaces.

Our choice of $U$=5 eV follows from our examination of the effect of
different $U$ values on both the geometry and the energy bands of rutile
TiO$_2$. While a larger $U$ value can give a wider band gap, it would
make a significant distortion in the atomic structure. Inclusion of Dudarev
$U$=5 eV term acting on Ti 3$d$ electrons gave reasonable values for the
atomic positions with bond lengths in agreement with the experimental data
for the stoichiometric, reduced, and the reconstructed rutile (110) surfaces.
This choice is also consistent with the previous calculations describing the
reduced~\cite{morgan,calzado,nolan} and reconstructed~\cite{celik} surfaces.
Spin polarization was also found to be important in determining the
semiconducting ground state of reconstructed surfaces and in correctly
describing Ti defect states while  it is insignificantly small for the
stoichiometric surface. Our PBE+$U$ calculations determined the ground
states with spin multiplicities of 2 and 6 per surface of (1$\times$2) unit
cells for Onishi and Park models, respectively.


\begin{table}[tbh]%
\caption{PBE+$U$ results of atomic positions for the Onishi model with
$U$=5 eV.\label{table1}}
\begin{ruledtabular}
\begin{tabular}{cccc|ccc}
& \multicolumn{3}{c}{Theoretical ({\AA})} &
\multicolumn{3}{c}{Experimental\footnote{LEED data in Ref.\cite{blancorey1}.}
({\AA})}
\\ \cline{2-7}
Atom & $x$~[001] & $y$~[1\={1}0] & $z$~[110] & $x$ & $y$ & $z$ \\[1mm] \hline
Ti(a) & ~~1.48 & ~~1.76 & $-5.77$ & 1.48 & 1.77 & $-5.99\pm$0.03 \\
Ti(b) & ~~1.48 &$-0.14$ & $-3.36$ & 1.48 & 0.00 & $-3.14\pm$0.07 \\
Ti(c) &$-0.04$ & ~~3.23 & $-3.28$ & 0.00 & 3.28 & $-3.27\pm$0.06 \\
Ti(d) & ~~1.48 & ~~6.44 & $-3.03$ & 1.48 & 6.49 & $-3.08\pm$0.05 \\
O(1)  &$-0.04$ & ~~2.08 & $-6.81$ & 0.00 & 1.99 & $-7.16\pm$0.24 \\
O(2)  & ~~1.48 & $-0.15$& $-5.53$ & 1.48 & 0.00 & $-5.23\pm$0.07 \\
O(3)  & ~~1.48 & ~~3.35 & $-4.62$ & 1.48 & 3.07 & $-4.60\pm$0.11 \\
O(4)  &$-0.04$ & ~~1.30 & $-3.86$ & 0.00 & 1.25 & $-3.21\pm$0.12 \\
O(5)  &$-0.04$ & ~~5.20 & $-3.35$ & 0.00 & 5.22 & $-3.54\pm$0.06 \\
O(6)  & ~~1.48 & $-0.13$ & $-1.35$ & 1.48 & 0.00 & $-1.30\pm$0.22 \\
O(7)  & ~~1.48 & ~~3.09 & $-2.01$ & 1.48 & 3.28 & $-2.03\pm$0.22 \\
O(8)  & ~~1.48 & ~~6.44 & $-1.23$ & 1.48 & 6.49 & $-1.31\pm$0.12
\end{tabular}
\end{ruledtabular}
\end{table}
\begin{table}[htb]%
\caption{PBE+$U$ results of atomic positions for the Park model
with $U$=5 eV.\label{table2}}
\begin{ruledtabular}
\begin{tabular}{cccc|ccc}
& \multicolumn{3}{c}{Theoretical ({\AA})} &
\multicolumn{3}{c}{Experimental\footnote{LEED data in Ref.\cite{blancorey2}.}
({\AA})}
\\ \cline{2-7}
Atom & $x$~[001] & $y$~[1\={1}0] & $z$~[110] & $x$ & $y$ & $z$ \\[1mm] \hline
Ti(a) & ~~0.02 & 1.56 & $-5.48$ & 0.00 & 1.57 & $-5.12\pm$0.14 \\
Ti(b) & ~~1.48 & 0.00 & $-2.80$ & 1.48 & 0.00 & $-3.14\pm$0.10 \\
Ti(c) & $-0.05$& 3.34 & $-3.15$ & 0.00 & 3.32 & $-3.26\pm$0.10 \\
Ti(d) & ~~1.48 & 6.58 & $-3.12$ & 1.48 & 6.49 & $-3.52\pm$0.06 \\
O(1)  & ~~1.65 & 0.00 & $-5.89$ & 1.48 & 0.00 & $-5.30\pm$0.12 \\
O(2)  & ~~1.47 & 2.83 & $-5.12$ & 1.48 & 3.14 & $-4.63\pm$0.14 \\
O(3)  & $-0.04$& 1.28 & $-3.45$ & 0.00 & 1.26 & $-3.32\pm$0.18 \\
O(4)  & $-0.04$& 5.32 & $-3.57$ & 0.00 & 5.22 & $-3.48\pm$0.16 \\
O(5)  & ~~1.48 & 0.00 & $-0.90$ & 1.48 & 0.00 & $-1.38\pm$0.32 \\
O(6)  & ~~1.47 & 3.24 & $-1.85$ & 1.48 & 3.24 & $-2.04\pm$0.12 \\
O(7)  & ~~1.48 & 6.58 & $-1.32$ & 1.48 & 6.49 & $-1.28\pm$0.20
\end{tabular}
\end{ruledtabular}
\end{table}

\begin{figure}[tbh]
\epsfig{file=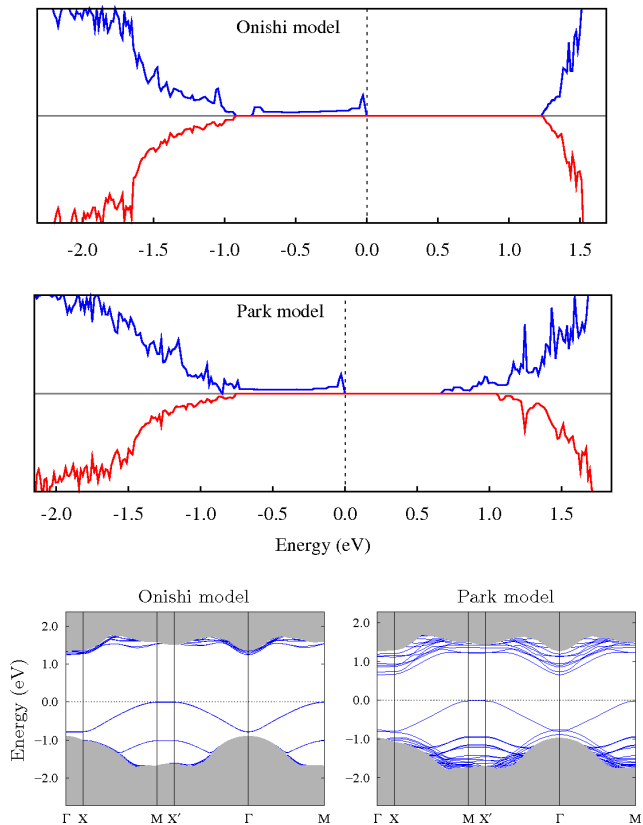,width=8.4cm}
\caption{Density of states (DOS) plots and energy band diagrams for the
majority electrons belonging to Onishi~\cite{onishi} and
Park~\cite{park} models of TiO$_2$(110)-(1$\times$2) surface.
\label{fig2}}
\end{figure}

\begin{figure}[tbh]
\epsfig{file=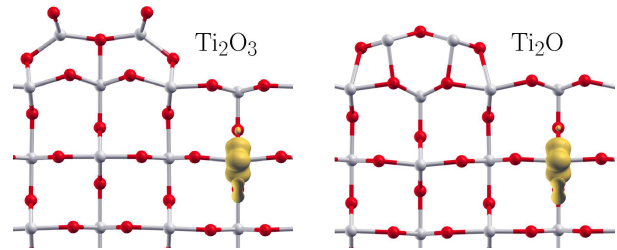,width=8.0cm}
\caption{Partial charge densities of the gap states.
\label{fig3}}
\end{figure}

Calculated atomic coordinates and the LEED results~\cite{blancorey1,blancorey2}
are only slightly different as shown in Table~\ref{table1} and
Table~\ref{table2} for each of the model structures. The most
noticeable deviation is seen in the $z$ component of O(4) in the Onishi
model. While keeping distortion to the atomic positions small, PBE+$U$ with
$U$=5 eV reproduces the gap states as shown in Fig.~\ref{fig2}. Defect states
has been found to be 1.24 eV and 0.68 eV below the CB for Onishi and Park
models, respectively, in agreement with experiments~\cite{henrich,henderson}. In
the Onishi model, the gap state with a dispersion width of 0.65 eV reveal charge
delocalization to the interstitial Ti atom just below in-plane Ti5c showing
$d_{z^2}$ character and to the oxygen atom beneath it as shown in
Fig.~\ref{fig3}.

As finding the almost correct position of the experimentally observed Ti $3d$
states, describing the origin of those states is also important. According to
the ultraviolet photoelectron spectroscopy (UPS) and LEED experiments, the
excess electrons upon O loss are believed to delocalize around
the surface Ti atoms, adjacent to the vacancies~\cite{henrich}.
Although this vacancy model is supported by some calculations \cite{valentin,
morgan}, there are studies that indicate the role of subsurface Ti atoms about
the delocalization of the excess charge~\cite{onishi,bennet2,wendt}.
Since pure DFT gives a clean band gap, based on their STM and UPS measurements
Wendt~\textit{et al.} proposed that the gap states originate from Ti atoms
diffused into interstitial sites, not from the surface Ti atoms adjacent to
bridging O vacancies~\cite{wendt}. Without a need to such an additional
interstitial Ti atom, delocalization of the excess charge to the subsurface
Ti, responsible for the gap state, emerges from DFT+$U$ calculations as shown in
Fig.~\ref{fig3}.

Surface free energy is a good measure to compare the thermodynamic stability of
model row formations. Stoichiometric cell is a stack of Ti$_4$O$_8$-(1$\times$2)
units that act as bulk layers around the central regions as taken into account
in Ref.~\cite{elliot2}. Our ten Ti-layer slab model can be expressed as
Ti$_n$O$_{2n}$ with $n=40$. Reduced reconstructions are led by surface O removal
from the stoichiometric surface. For our symmetrical slab cell, Ti$_n$O$_{2n-m}$
represents both the Onishi ($m$=2) and the Park ($m$=6) models. Then, the
surface energies were calculated by the relation,
\[
\sigma=\frac{1}{2A}\;(E_n^{\rm slab}-nE^{\rm bulk}+mE^{\rm O})\, ,
\]
where $A$ is the surface unit cell area, $E_n^{\rm slab}$, $E^{\rm bulk}$, and
$E^{\rm O}$ are the energies of a Ti$_n$O$_{\rm 2n-m}$ slab, of a bulk
Ti$_4$O$_8$ unit ($-89.3$ eV) and of an oxygen atom in its molecular form,
respectively. Division by two is because the slab has two reconstructed surfaces
on its both faces. In an experiment, the surface layer is an interface between
O$_2$ gas phase and TiO$_2$ bulk crystal. Thermal equilibrium can exist if the
chemical potential of the atomic species are equal in all these phases that
come into contact with each other. Therefore, determination of the chemical
potential of oxygen atom, $E^{\rm O}$, limits the accuracy of the calculated
surface energies. We found the binding energy of an O$_2$ molecule to be 6.07 eV
which is significantly larger than the experimental value of 5.26 eV~\cite{nist}.
The tendency of DFT to overestimate it, was also reported
previously~\cite{morgan,kowalski}. Therefore, in order to obtain a more
reasonable energy value for $E^{\rm O}$, we adopted the experimental
binding energy of O$_2$ (5.26 eV) and DFT result of an isolated O atom ($-1.89$
eV). It gives $-4.52$ eV for $E^{\rm O}$. By using this reference chemical
potential we calculated the formation energy of bulk TiO$_2$ from metallic bulk
Ti and O$_2$ molecule as 9.72 eV, in excellent agreement with the experimental
value of 9.73 eV~\cite{nist}. Therefore thermodynamic equilibrium between the
surface and the bulk crystal can also be reached.

\begin{figure}[tbh]
\epsfig{file=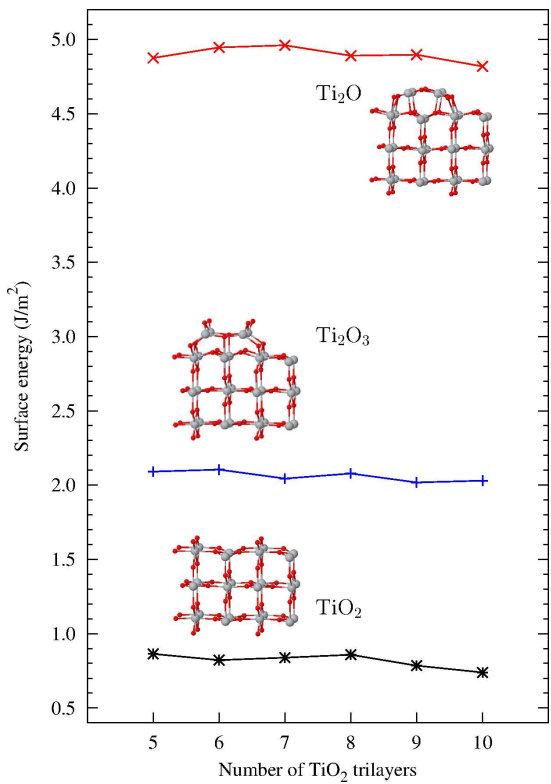,width=7.2cm}
\caption{Calculated free energies of TiO$_2$(110) surface models.\label{fig4}}
\end{figure}

In Fig.~\ref{fig4} we compare relative stabilities of stoichiometric and
reconstructed slabs through their surface energies with varying number of
Ti-layers. Surface energetics of the stoichiometric slab at the Hartree--Fock and
at the standard DFT levels were reported to exhibit odd--even oscillations with
the number of layers~\cite{bates,kiejna,labat,kowalski}. Our results show that
the oscillation of both the stoichiometric and reduced surface energies with
slab thickness settles down by the inclusion of $U$=5 (Fig.~\ref{fig4}). Using
ten Ti-layer cells we calculated well converged surface free energies as 0.74,
2.03, and 4.82 J$\cdot$m$^{-2}$, for the stoichiometric and reconstructed
(Onishi and Park models) surfaces, respectively. Our results are significantly
different from the calculations carried out with pure functionals. For instance,
Morgan~\textit{et al.} calculated the surface energy of stoichiometric case to
be 0.58 J$\cdot$m$^{-2}$ using GGA and to be 0.83 J$\cdot$m$^{-2}$ using GGA+$U$
($U$=4.2 eV) over (4$\times$2) supercell with five Ti-layers~\cite{morgan}. The
latter is in good agreement with our GGA+$U$ value of 0.86 J$\cdot$m$^{-2}$
calculated with 5-Ti-layer stoichiometric (1$\times$2) supercell. For the
Ti$_2$O$_3$ added row model, Elliot~\textit{et al.} found the surface energy
using spin-polarized DFT to be 3.29$\pm$0.08 J$\cdot$m$^{-2}$ which is corrected
by the absolute energy value of an isolated oxygen atom~\cite{elliot1}. On the
other hand, for recent Ti$_2$O model of Park~\textit{et al.}, no calculations
for the surface energy were reported.

Relaxed bulk termination appears to be the most stable surface. Once the
TiO$_2$(110) surface is reduced (followed by a reconstruction), O$_2$ exposure
can not restore the stoichiometric form again~\cite{wendt}. Thus, comparison
between the reconstruction models is more meaningful. Formation energy of
the Ti$_2$O$_3$ added row proposed by Onishi and Iwasawa (with a ground state
spin polarization of $\mu=2$) is only 1.29 J$\cdot$m$^{-2}$ higher relative to
that of the stoichiometric surface. 

Standard DFT incorrectly predicts Ti $3d$ excess charge to occupy the bottom
of the CB leading to metallization for the reduced and reconstructed surfaces.
PBE+$U$ method with $U$=5 reproduces experimentally observed gap states for the
(1$\times$2) reconstructions as well as for the oxygen vacancies on the rutile
(110) surfaces. Finally, inclusion of a suitably chosen $U$ parameter in the
calculations for the reconstructed rutile TiO$_2$(110) surface, is a simple and
promising way of restoring its semiconducting nature by reproducing the band-gap
states which arise from delocalization of Ti $3d$ excess charge to subsurface Ti
sites. According to their surface energies, Onishi's added row model is more
stable than the Park model. Therefore, Ti$_2$O$_3$ added row model confirms
existing experimental observations and can still be assigned as the (1$\times$2)
long range order on the rutile (110) surface.

\begin{acknowledgments}
This work was supported by T\"{U}B\.{I}TAK, The Scientific and Technological
Research Council of Turkey (Grant \#110T394). Computational resources were
provided by ULAKB\.{I}M, Turkish Academic Network and Information Center.
\end{acknowledgments}

\end{document}